\documentclass[twocolumn]{aastex63}

\DeclareRobustCommand{\ion}[2]{%
\relax\ifmmode
\ifx\testbx\f@series
{\mathbf{#1\,\mathsc{#2}}}\else
{\mathrm{#1\,\mathsc{#2}}}\fi
\else\textup{#1\,{\mdseries\textsc{#2}}}%
\fi}

\newenvironment{myfont}{\fontfamily{lmtt}\selectfont}{\par}
\DeclareTextFontCommand{\textmyfont}{\myfont}

\usepackage{amsmath,amssymb}
\usepackage{natbib}
\usepackage[normalem]{ulem}
\usepackage[T1]{fontenc}
\usepackage{soul,xcolor}

\def\n{$\rm{n_{H}}$}
\def\u{$\rm{U}$}
\def\feseven{\rm{[Fe \sc{vii}]}}
\def\sisix{\rm{[Si \sc{vi}]}}
\def\siten{\rm{[Si \sc{x}]}}
\def\seight{\rm{[S \sc{viii}]}}

\def\LLEdd{$L\mathrm{_{bol}}/L\mathrm{_{Edd}}$}

\received{June 1, 2019}
\revised{January 10, 2019}
\accepted{\today}
\submitjournal{ApJ}

\shorttitle{Probing accretion in AGN}
\shortauthors{Rodr\'{\i}guez-Ardila et al.}
\graphicspath{{./}{figures/}}

\begin{document}

\setstcolor{red}

\title{A novel black-hole mass scaling relation based on Coronal lines and supported by accretion predictions  }

\correspondingauthor{Alberto Rodr\'{\i}guez-Ardila}
\email{aardila@lna.br}

\author[0000-0002-7608-6109]{Alberto Rodr\'{\i}guez-Ardila}
\affiliation{LNA/MCTIC, Rua dos Estados Unidos, 154. Bairro das Na\c c\~oes, Itajub\'a, MG 37501-591, Brazil}
\affiliation{Divis\~ao de Astrof\'{\i}sica, INPE, Avenida dos Astronautas 1758, S\~ao Jos\'e dos Campos, 12227-010, SP, Brazil}

\author[0000-0002-3585-2639]{Almudena Prieto}
\affiliation{Instituto de Astrof\'{\i}sica de Canarias and Universidad de La Laguna, V\'{\i}a L\'actea s/n, E-38205, Tenerife, Spain} 

\author[0000-0002-5854-7426]{Swayamtrupta Panda}
\affiliation{Center for Theoretical Physics, Polish Academy of Sciences, Al. Lotnik{\'o}w 32/46, 02-668 Warsaw, Poland}
\affiliation{Nicolaus Copernicus Astronomical Center, Polish Academy of Sciences, ul. Bartycka 18, 00-716 Warsaw, Poland}

\author[0000-0001-9719-4523]{Murilo Marinello} 
\affiliation{LNA/MCTIC, Rua dos Estados Unidos, 154. Bairro das Na\c c\~oes, Itajub\'a, MG 37501-591, Brazil}









\begin{abstract}

Getting insights on the shape and nature of the ionizing continuum in astronomical objects is often done via indirect methods as high energy photons are absorbed by our Galaxy. This work explores the ionization continuum of  active galactic nuclei (AGN) using the ubiquitous coronal lines. Using \textit{bona-fide} BH mass estimates from reverberation mapping and the line ratio [\ion{Si}{vi}]~1.963$\micron$/Br$\gamma_{\rm broad}$ as  tracer of the AGN ionizing continuum, a novel BH-mass scaling relation of the form log($M_{\rm BH}) = (6.40\pm 0.17) - (1.99\pm 0.37) \times$ log ([\ion{Si}{vi}]/Br$\gamma_{\rm broad})$, over the BH mass interval, $10^6 - 10^8$ M$_{\odot}$ with dispersion 0.47 dex is found. Following on the thin accretion disc approximation and after surveying a basic parameter space for coronal lines (CL) production, we believe that a key parameter driving this anti-correlation is the effective temperature of the accretion disc, this being effectively sampled by the coronal line gas. Accordingly, the  observed  anti-correlation becomes formally in line with the thin accretion disc prediction $T_{\rm{disc}} \propto {M_{\rm BH}}^{-1/4}$.

\end{abstract}

\keywords{accretion, accretion disks -- radiative transfer -- techniques: spectroscopic -- galaxies: active -- quasars: emission lines }


\section{Introduction} \label{sec:intro}
The ionizing continuum of active galactic nuclei (AGN) is for most cases a non-direct observable as high energy photons, namely in the  10 eV -- 400 eV range, are absorbed by neutral HI of our Galaxy. These photons are however responsible for 99\% of all emission lines observed in the ultraviolet (UV), optical and near-infrared  spectrum.

It is often assumed  that the bulk of this energy is  produced in the atmosphere of a  geometrically thin, optically thick accretion disc (AD) around a supermassive black hole \citep{lynden-bell69, ss73, shields78}.
Observations of quasars and photoionization equilibrium models suggest the ionizing continuum to peak in the UV, its  shape being close to a black-body  spectrum whose peak temperature is taken as the disc's effective temperature as it is around this temperature that most of the energy is radiated \citep[e.g.][]{frank+02}. Still, the diversity of AGN line spectra can equally be  reproduced by an otherwise UV power-law spectrum, or combination of power laws,  to account for the ionizing continuum \citep{dn79, fo86, ferlandetal2020}.

Photoionization models are the common tool to infer on the shape and nature of the ionizing continuum. Yet, observations restrict in most cases their use to the brightest emission lines from the UV to the near-IR, which probe only the energy range between 12 to 54 eV. Because of their high ionization potential (IP) above 100~eV,  coronal lines (CL) are optimal features to  map the shape of the ionizing continuum above the $\sim 54 $~eV  threshold \citep[][and references therein]{binetteetal1996, ferguson+97, prieto+00, contini02}. CL spread over the X-rays, optical and IR spectrum. Although often fainter than the classical medium-ionzation lines used for photoionization diagnosis, high angular resolution in  nearby AGN has delivered extremely high signal-to-noise measurements of CLs in the optical, and particularly in the near-IR where are   often the most conspicuous features \citep[e.g.][]{reunanen03, prieto+05, muller-sanchez+11, rodriguez+17}.

Here, we make use of a set of accurate measurements of optical and near-IR CL to examine the shape of the ionizing continuum in the 100 - 400 eV range, and in turn, to test the  accretion theory prediction on the dependence of the black hole (BH) mass with the effective temperature of the accretion disc $T_{\rm{disc}}$, as $T_{\rm{disc}} \propto \rm{M_{BH}}^{-1/4}$.
Following accretion theory, for similar accretion rates, high BH  masses are expected to have discs with a larger inner radius, and thus become cooler, i.e. be characterised by a lower effective temperature $T_{\rm{disc}}$, than their lower mass counterparts. Accordingly, discs of  BH mass of  $\sim 10^{7 - 8}$~M$_{\odot}$ are expected to peak in the extreme UV, which would correspond to what is often referred to as the  Big Blue Bump \citep[BBB,][]{czerny87} - e.g  in 3C 273 this is $\sim  $10 eV \citep{2010MNRAS.402..724P} -  whereas those of lower BH mass, $< 10^6$~M$_{\odot}$ are expected to peak towards higher energies $\geq$ 100~eV \citep[e.g.][]{frank+02, cann+18}. The change in the disc temperature across this BH mass interval is at most a factor seven (see below).

The work presents a novel approach to estimate BH masses as a function of specific  coronal lines normalised to H broad line emission, so far for Type~I AGN only. Possible explanation for the observed BH mass - CL relations are examined in the context of the predicted dependence between accretion disc temperature and BH mass.

\section{Data selection}

Objects in this work are selected by having BH masses determined by reverberation mapping and single epoch  optical and/or near-IR spectra with accurate  CL measurements.  The first criterion restricts the sample to Type~I sources only. The second avoids variability issues. Although we give preference to sources with both optical and near-IR spectra available, this final criterion  could not always be fulfilled.  

The CL used are [\ion{Fe}{vii}]~$\lambda$6087~\AA\ in the optical and [S\,{\sc viii}]~0.991~$\mu$m, [Si\,{\sc x}]~1.432~$\mu$m and [Si\,{\sc vi}]~1.964~$\mu$m in the near-IR. They are among the strongest  CL in AGN  \citep{reunanen03, rodriguez+11, lamperti+17} and span a wide IP range, 100 - 351~eV. In addition,  \ion{H}{1} lines of H$\beta$, Pa$\beta$ and Br$\gamma$ are employed. The whole set samples the ionizing continuum over the 13.6 - 351~eV range.  Near-IR CL were preferred because of their reduced extinction. Optical CL [\ion{Fe}{vii}] was also selected because of its strength, still moderate extinction,  and IP close to that of [Si\,{\sc vi}].

The final working sample of objects has 31 AGN (Table~\ref{tab:data}). For a subsample, the pertinent optical and near-IR line are presented in this work for the first time.  
For all other sources, line ratios or spectra already described in other publications were employed (see below). BH masses are from \citet{bentz+15} compilation.

\subsection{Optical Spectroscopy}

Optical spectra were taken from a variety of sources, as indicated in the last column of Table~\ref{tab:data}. In more than half of the sample, spectra from the Sloan Digital Sky Survey (SDSS) data release~7 \citep{abazajian/2009} were employed. SDSS delivers fully wavelength and flux calibrated spectra. Therefore, data reduction for these objects will not be discussed here. Similarly, archival flux-calibrated spectra for Ark\,564 taken by the Faint Object Spectrograph (FOS) on-board the Hubble Space Telescope (HST) were employed. Details of observations and reduction of this target will not be addressed here.  NGC\,4051 employs archival spectroscopy available from the Nasa Extragalactic Database (NED). Details of that observation can be found in \citet{moustakas+06}. Mrk~335 was observed using the 2.15~m telescope at the Complejo Astron\'omico El Leoncito (CASLEO). Details of the observations and data reduction can be found in  \citet{rodriguez+02}. Spectra for Fairall~9, NGC~4151, and Mrk~509 were extracted from the AGN Watch Project\footnote{http://www.astronomy.ohio-state.edu/~agnwatch/}. These data is fully reduced and details of that procedure can be found elwhere.

The second major source of optical data is the 4.1 m Southern Observatory for Astrophysical Research (SOAR) Telescope at Cerro Pachon, Chile. The observations were carried out using the Goodman Spectrograph \citep{clemens/2004}, equipped with a 600~l/mm grating and a 0.8 arcsec slit width, giving a resolution R$\sim$1500. 
In addition to the science frames, standard stars \citep{baldwin/1984} were observed for flux calibration. HgAr arc lamps were taken after the science frames for wavelength calibration. Daytime calibrations include bias and flat field images.
 
The data were reduced using standard {\sc iraf} tasks. It includes subtraction of the bias level and division of the science and flux standard star frames by a  normalized master flat-field image. Thereafter, the spectra were wavelength calibrated by applying 	the dispersion solution obtained from the arc lamp frames. 
Finally, the spectra of standard stars were extracted and combined to derived the sensitivity function, later applied to the 1D science spectra. The final products are wavelength and flux calibrated optical spectra.

In all cases above,  the final spectra were corrected for Galactic extinction using the extinction maps of \citet{schlafly/2011} and the \citet{cardelli/1989} extinction law. Figures~\ref{fig:optspec1} and~\ref{fig:optspec2} show the optical sample in the spectral regions centred around the H$\beta$ and [\ion{Fe}{vii}]~$\lambda$6087 lines.

\subsection{NIR Spectroscopy}

Most of the NIR emission line flux ratios employed in this work were extracted from \citet{riffel+06}. For targets not reported in that publication, observations were obtained using either The Gemini Near-Infrared Spectrograph (GNIRS) attached to the Gemini North Telescope or the the ARCOiRIS spectrograph, mounted on either Blanco or SOAR Telescopes. Note that ARCOiRIS was installed on Blanco since 2017 and up to 2019, when it was then moved to SOAR with no modifications regarding their setup. Below we describe the observations and data reduction procedures, noting that no distinction between Blanco and SOAR is made. Both data collection and treatment is made employing the same observing strategy and reduction pipelines.

\subsubsection{ARCOiRIS Blanco/SOAR data}

NIR spectra of Fairall~9, 3C~120, Mrk~707, NGC\,3783, Mrk\,1310, NGC\,841, NGC\,6814 and NGC\,7469 were obtained using the ARCoIRIS spectrograph attached to either the 4\,m Blanco Telescope or the Southern Observatory for Astrophysical Research 4.1 m telescope atop Cerro Pach\'on, Chile. The science detector employed is a 2048 $\times$ 2048 Hawaii-2RG HgCdTe array with a sampling of 0.41 arcsec/pixel. The slit assembly is 1.1 arcsec wide and 28 arcsec long. The delivered spectral resolution R is $\sim$3500 across the six science orders. Observations were done nodding in two positions along the slit. Right before or after the science target, a telluric star, close in airmass to the former, was observed to remove telluric features and to perform the flux calibration. CuHgAr frames were also observed at the same position as the galaxies for wavelength calibration.

The spectral reduction, extraction and wavelength calibration procedures were performed using {\sc spextool v4.1}, an IDL-based software developed and provided by the SpeX team \citep{cushing/2004}  with some modifications specifically designed for the data format and characteristics of ARCoIRIS, written by Dr. Katelyn Allers (private communication). Telluric features removal and flux calibration were done using {\sc xtellcor} \citep{vacca/2003}. The different orders were merged into a single 1D spectrum from 1 to 2.4~$\mu$m using the {\sc xmergeorders} routine. We then corrected these data for Galactic extinction using the \citet{cardelli/1989} law and the extinction maps of \citet{schlafly/2011}.

\subsubsection{GNIRS/Gemini spectroscopy}

Near-infrared spectra of NGC\,4395 and Ark\,564 were collected using
The Gemini Near-IR spectrograph \citep[GNIRS,][]{elias/2006} in the cross-dispersed mode. It allows simultaneous z+J, H and K band observations, covering the spectral range 0.8\,$-$\,2.5$\mu$m in a single exposure.
GNIRS science detector consist of an ALADDIN 1k $\times$ 1k InSb array. The instrument setup includes a 32~l/mm grating and a 0.8$\times7$ arcsec slit, giving a spectral resolution of R$\sim$1300 (or 320~km\,s$^{-1}$ FWHM). Individual exposures were taken, nodding the source in a ABBA pattern along the slit.
Right after the observation of the science frames, an A0V star was observed at a similar airmass, with the purpose of flux calibration and telluric correction.
 
The NIR data were reduced using the XDGNIRS pipeline (v2.0)\footnote{Based on the Gemini IRAF packages}, which delivers a full reduced, wavelength and flux calibrated, 1D spectrum with all orders combined \citep{mason/2015}. 
Briefly, the pipeline cleans the 2D images from radiative events and prepares a master flat constructed from quartz IR lamps to remove pixel to pixel variation. 
Thereafter, the s-distortion solution is obtained from daytime pinholes flats and applied to the science and telluric images to rectify them. 
Argon lamp images are then used to find the wavelength dispersion solution, followed by the extraction of 1D spectra from the combined individual exposures. 
The telluric features from the science spectrum are removed using the spectrum of a A0V star. 
Finally, the flux calibration is achieved assuming a black body shape for the  standard star \citep{pecaut/2013} scaled to its $K$-band magnitude \citep{skrutskie/2006}. 
The different orders are combined in to a single 1D spectrum and  corrected for Galactic extinction using the \citet{cardelli/1989} law and the extinction maps of \citet{schlafly/2011}.  

The final reduced NIR spectra in the regions of interest to this work are shown in Fig. \ref{fig:nirspec1}. 

In order to measure the flux of the lines for the subsample of objects described above, we modelled the observed profiles with a suitable function that best represents them and then integrated the flux under that function. To this purpose we employ the {\sc liner} routine (\citealt{pogge/1993}).  This software performs a least-square fit of a model line profile (Gaussian, Lorentzian, or Voight functions) to a given line or set of blended lines to determine the flux, peak position and FWHM of the individual components. Typically, one or two Gaussian components were necessary to represent the coronal lines. For the permitted lines of \ion{H}{i} a broad component associated to the BLR was employed. In this process, the underlying continuum emission was approximated by a linear fit. 

For the optical part, the measurement of the H$\beta$ flux was preceded by the removal of the underlying power-law continuum and the pseudo-continuum produced by the \ion{Fe}{ii} lines that contaminates H$\beta$. This was done following the prescription of \citet{bg92}.

Table~\ref{tab:data} shows the measured optical emission line flux ratio between [\ion{Fe}{vii}]~$\lambda$6087 and the broad component of H$\beta$ (column~3) and the  NIR ratios for [\ion{Si}{vi}]~1.964~$\mu$m/Br$\gamma$ (column 4), [\ion{Si}{x}]~1.431~$\mu$m/Pa$\beta$ (column 5), and [\ion{S}{viii}]~0.9914~$\mu$m/Pa$\beta$ (column 6). For the latter three ratios,  the flux associated to the broad component of the Brackett and Paschen line were employed. Table~\ref{tab:fluxes} in the Appendix list the flux of the broad components of H$\beta$, Pa$\beta$ and Br$\gamma$ for the AGN with available NIR data. Note that because the optical and NIR spectra were taken on different dates and in most cases, different telescopes, the intrinsic line ratios H$\beta$/Pa$\beta$ and H$\beta$/Br$\gamma$ largely depart from their theoretical value. This, however, does not affect our results as we do not employ in our analysis line ratios between the two spectral regions.

\begin{table*}[]
\centering
\caption{Black hole mass and CL ratios for the galaxy sample.}
\label{tab:data}
\resizebox{17cm}{!}{%
\hspace*{-2cm}\begin{tabular}{lcccccc}
\hline\hline\noalign{\vskip 0.1cm}
\multicolumn{1}{l}{Galaxy} & log $M_{\rm BH}^1$ & [\ion{Fe}{vii}]/H$\beta ^{2}$ & [\ion{Si}{vi}]/Br$\gamma^{3}$ & [\ion{Si}{x}]/Pa$\beta$ & [\ion{S}{viii}]/Pa$\beta$ & Data Source \\

\hline\noalign{\vskip 0.1cm}
Mrk\,335 & 7.23$\pm$0.04 & 0.048$\pm$0.005 & 0.40$\pm0.09^{4}$ & 0.04$\pm$0.01 & 0.022$\pm$0.006 & 6 \\
Fairall$\sim$9 & 8.29$\pm$0.09 & 0.038$\pm$0.003 & 0.11$\pm0.02^{4}$ & 0.08$\pm$0.01 & 0.04$\pm$0.01 & 5,10 \\
NGC\,863 & 7.57$\pm$0.06 & ... & 0.24$\pm$0.07 & ... & ... & 6 \\
3C\,120 & 7.74$\pm$0.04 & ... & 0.33$\pm0.06^{4}$ & 0.08$\pm$0.02 & 0.03$\pm$0.01 & 5 \\
Mrk\,707 & 6.50$\pm$0.10$^a$ & 0.025$\pm$0.002 & 0.38$\pm0.04^{4}$ & ... & ... & 5,7 \\
Mrk\,110 & 7.29$\pm$0.10 & 0.05$\pm$0.002 & ... & ... & ... & 7 \\
NGC\,3227 & 6.78$\pm$0.10 & ... & 0.75$\pm$0.20 & ... & 0.012$\pm$0.004 & 3 \\
Mrk\,142 & 6.29$\pm$0.10 & 0.02$\pm$0.004 & ... & ... & ... & 7 \\
SBS\,1116+583A & 6.56$\pm$0.09 & 0.01$\pm$0.002 & ... & ... & ... & 7 \\
PG\,1126-041 & 8.08$^b$ & ... & 0.28$\pm$0.02 & 0.04$\pm$0.01 & 0.025$\pm$0.002 & 3 \\
NGC\,3783 & 7.37$\pm$0.08 & 0.05$\pm$0.002 & 0.42$\pm0.09^{4}$ & 0.05$\pm$0.01 & 0.023$\pm$0.003 & 5 \\
Mrk\,1310 & 6.21$\pm$0.08 & 0.03$\pm$0.002 & 0.57$\pm0.17^{4}$ & 0.06$\pm$0.01 & 0.06$\pm$0.01 & 5 \\
NGC\,4051 & 6.13$\pm$0.12 & 0.125$\pm$0.01 & 0.96$\pm$0.11 & 0.33$\pm$0.02 & 0.205$\pm$0.030 & 3,9 \\
NGC\,4151 & 7.55$\pm$0.05 & 0.02$\pm$0.001 & 0.51$\pm$0.05 & 0.05$\pm$0.01 & 0.057$\pm$0.003 & 3,10 \\
Mrk\,202 & 6.13$\pm$0.17 & 0.02$\pm$0.002 & ... & ... & ... & 7 \\
Mrk\,766 & 6.82$\pm$0.05 & 0.03$\pm$0.002$^d$ & 0.78$\pm$0.10 & 0.05$\pm$0.01 & 0.045$\pm$0.002 & 3 \\
Mrk\,50 & 7.42$\pm$0.06 & 0.005$\pm$0.001 & ... & ... & ... & 7 \\
NGC\,4395 & 5.45$\pm$0.13 & 0.09$\pm$0.005 & 1.18$\pm0.10^{4}$ & 0.02$\pm$0.01 & 0.053$\pm$0.006 & 7,8 \\
Mrk\,771 & 7.76$\pm$0.20 & 0.03$\pm$0.002 & ... & ... & ... & 7 \\
NGC\,4748 & 6.41$\pm$0.11 & ... & 0.93$\pm$0.06 & 0.06$\pm$0.02 & 0.137$\pm$0.042 & 3 \\
PG\,1307+085 & 8.54$\pm$0.13 & 0.01$\pm$0.002 & ... & ... & ... & 7 \\
MGC-6-30-15 & 6.60$\pm$0.12 & 0.017$\pm$0.002 & ... & ... & ... & 5 \\
NGC\,5548 & 7.72$\pm$0.02 & 0.04$\pm$0.003 & 0.61$\pm$0.09 & 0.11$\pm$0.01 & 0.122$\pm$0.011 & 3,7 \\
PG1448+273 & 6.97$\pm$0.08 & ... & 0.57$\pm$0.11 & ... & ... & 3 \\
Mrk\,290 & 7.28$\pm$0.02 & 0.02$\pm$0.002 & ... & ... & ... & 7 \\
Mrk\,841 & 8.10$^c$ & 0.008$\pm$0.002 & 0.20$\pm0.06^{4}$ & 0.02$\pm$0.01 & 0.024$\pm$0.002 & 5 \\
3C\,390.3 & 8.64$\pm$0.04 & 0.02$\pm$0.001 & ... & ... & ... & 7 \\
NGC\,6814 & 7.04$\pm$0.06 & ... & 0.13$\pm0.02^{4}$ & ... & ... & 5 \\
Mrk\,509 & 8.05$\pm$0.04 & ... & 0.17$\pm$0.02 & ... & ... & 3,10 \\
Ark\,564 & 6.59$\pm$0.17 & 0.06$\pm$0.007 & 1.08$\pm$0.10 & 0.30$\pm$0.01 & 0.101$\pm$0.006 & 8,11 \\
NGC\,7469 & 6.96$\pm$0.05 & 0.02$\pm$0.001 & 0.60$\pm$0.05 & 0.07$\pm$0.01 & 0.037$\pm$0.006 & 5 \\
\hline\\ 
\end{tabular}%
}
{
\begin{minipage}{0.9\textwidth}
 1. Masses are from \citet{bentz+15} unless stated otherwise; (\textit{a}) -- \citet{park+17}; (\textit{b}) -- \citet{dasyra+07}; (\textit{c}) -- \citet{woo+02}. 2. This optical emission line flux ratio was determined in this work unless otherwise stated. In all cases, the flux of the CL is normalized to the flux of the broad component of the \ion{H}{i} line. (\textit{d}) -- \citet{rodriguez+05}. 3. NIR emission line flux ratios from \citet{riffel+06} except when indicated. 4 -- This work. 5. -- SOAR; 6 -- CASLEO; 7 -- SDSS; 8 -- Gemini; 9 -- NED; 10 -- AGN Watch; 11 -- HST.    
\end{minipage}    

}
\end{table*}

\section{Coronal line diagnostic diagrams}
\label{sec3}
The dependence between the effective temperature of the disc and BH mass predicted by accretion theory \citep[cf.][]{frank+02} is investigated below using the CL emission as a proxy for the disc temperature. The approach was recently explored with photoionization simulations by \citet{cann+18} who show that for intermediate to low BH mass sources ($10^2 - 10^5$~M$_\odot$), lines of high IP are favoured with respect to those of lower IP. Here, we  expand the photoionization study to the high BH mass range, $10^6 - 10^9$~M$_\odot$, and confront predictions with observations. 

\begin{figure*}
	\includegraphics[width=\textwidth]{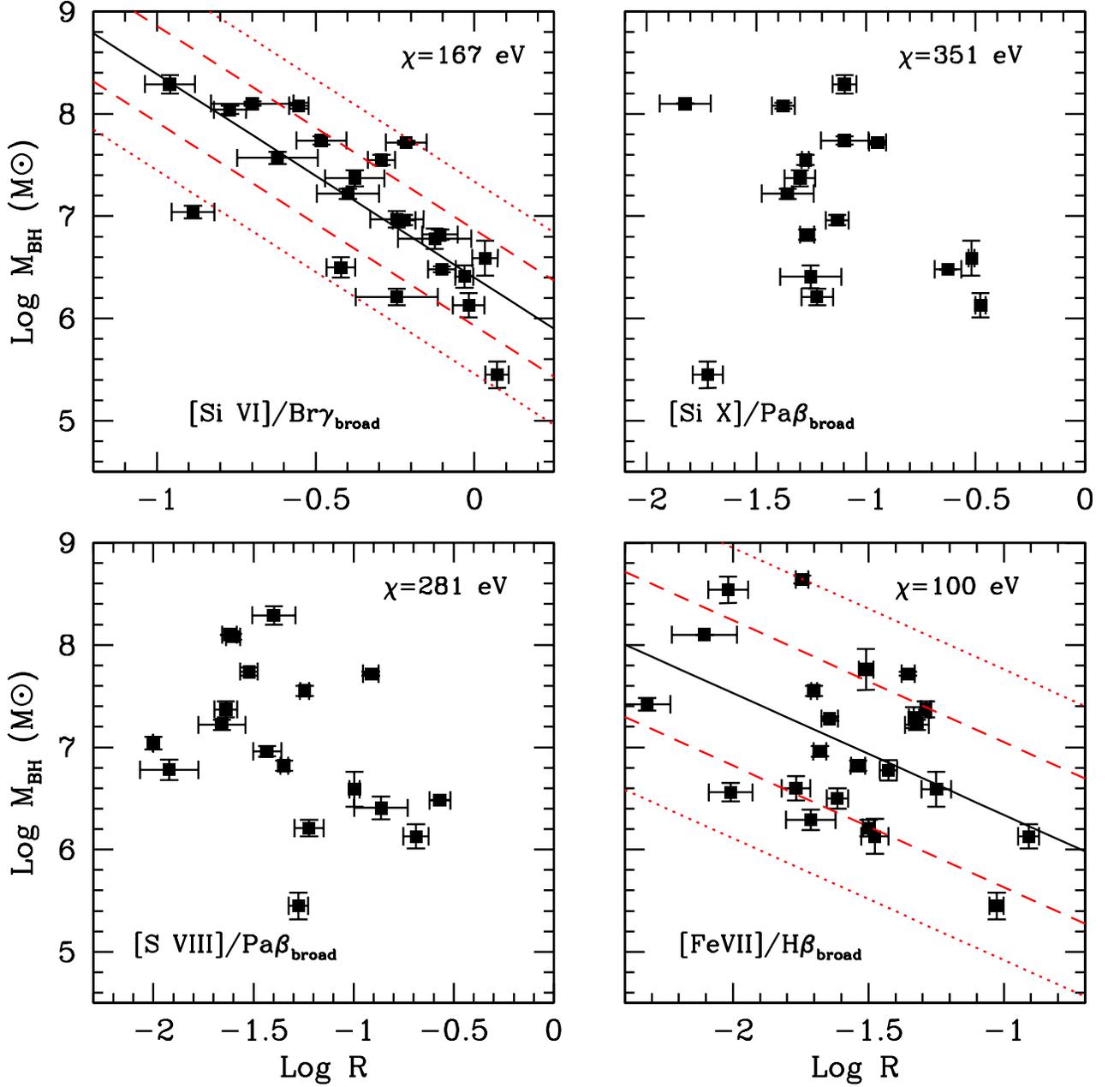}
    \caption{Observed CL emission normalised to the broad component of \ion{H}{I} versus black hole mass for the targets in this work.  The black  line is the best linear fit to the data and the red-dashed and -dotted lines show the 1$-\sigma$ and 2$-\sigma$ deviation, respectively.
    } 
    \label{fig:ratios}
\end{figure*}

Fig.~\ref{fig:ratios} presents new diagnostic  diagrams in which the BH mass for each object (see Column~2 of Table~\ref{tab:data}) in the sample is plotted against a specific CL flux normalised to the closest in wavelength \ion{H}{I} broad emission.  
The first diagram in Fig.~\ref{fig:ratios}-upper left panel, involving [\ion{Si}{vi}]~1.9641~$\mu$m/Br$\gamma_{\rm broad}$, IP [\ion{Si}{vi}] = 167 eV,  shows a clear linear trend between this ratio and    $M_{\rm BH}$  over almost  three orders of magnitude in BH mass. A linear regression yields  log($M_{\rm BH}) = (6.40\pm 0.17) - (1.99\pm 0.37) \times$ log ([\ion{Si}{vi}]/Br$\gamma_{\rm broad})$, and dispersion 0.47 dex (1 sigma) in BH mass.  The regression was carried out using the LtsFit package\footnote{\href{http://www-astro.physics.ox.ac.uk/~mxc/software/\#lts}{http://www-astro.physics.ox.ac.uk/mxc/software/lts}} \citep{capellari+13}, which accounts for the  errors in all variables. The Pearson correlation coefficient is $r$ = -0.76, with a null probability of Pr = 3.8$\times 10^{-5}$. 

A weak trend for the ratio involving [\ion{Fe}{vii}] (r=-0.5), and no trend for the IR CL with IP $>$ 260~eV, are found. The correlation index for [\ion{S}{viii}]/Pa$\beta_{\rm broad}$ is $r$ = -0.44, that for [\ion{Si}{x}]/Pa$\beta_{\rm broad}$, is $r$= -0.3.

In doing these diagrams, attention is paid to the following issues:  line  ratios are close   in wavelength to minimize reddening; the bulk of CL is produced at the inner parsecs and is of photoionisation origin  \citep[e.g.][]{prieto+00,prieto02,muller-sanchez+11,rodriguez-ardila06}, therefore the choice of  normalization to \ion{H}{i} broad is aimed at producing a most genuine tracer of the conditions next to the accretion disc;  the CL high critical density, $n_{\rm e}> 10^8$ cm$^{-3}$ warrants its survival in the inner high density environment of AGN. 
 
The CL in the diagrams, though,  are sensitive to different energy ranges of the ionizing continuum. This is illustrated in Fig~\ref{fig:continuum}, which shows a common used   parametrization of an AGN ionizing continuum  by a  Shakura-Sunyaev (SS) accretion disc \citep{ss73} that accounts for the big blue bump, and a combination of power-laws to account for the high energy range \citep[e.g.][]{osterbrock-ferland06,panda18b}.

\begin{equation}
    F_{\nu} = \nu^{\alpha_{uv}}\exp{\left(\frac{-h\nu}{kT_{disc}}\right)}\exp{\left(\frac{-kT_{IR}}{h\nu}\right)} + a\nu^{\alpha_x}
\label{eq:sed}    
\end{equation}
 
The first term is the parametrization of a SS disc, represented by an exponential cutoff at the disc effective temperature, $T_{\rm{disc}}$, and a power law with $\alpha_{\rm uv}$=-0.33 accounting for the low energy tail of the  disc. The low energy limit to the disk is set by the secondary IR-exponential cutoff at 0.01 Ryd. The high energies are represented by a broken power law, $\alpha_{\rm x}$= -1 and a high high-energy cutoff at 100 keV.

The IP  of the  lines used in this work are marked on the ionizing continuum in  Fig.~\ref{fig:continuum}. They  sample the bulk of the ionizing spectrum over the 13.6 - 351 eV energy range. The figure shows different ionizing continua for different $T_{\rm{disc}}$ and BH spin (see Sec. \ref{accretion-disc}). The increase in $T_{\rm disc}$ and spin leads to a progressive shift of the disc emission peak  towards higher energies. It can be observed that only when  spin is considered, the accretion disc samples adequately  the required high energies  to produced these CL.

\section{Coronal  emission as diagnostic of  accretion disc temperature}
\label{accretion-disc}

\begin{figure}
	\includegraphics[width=\columnwidth]{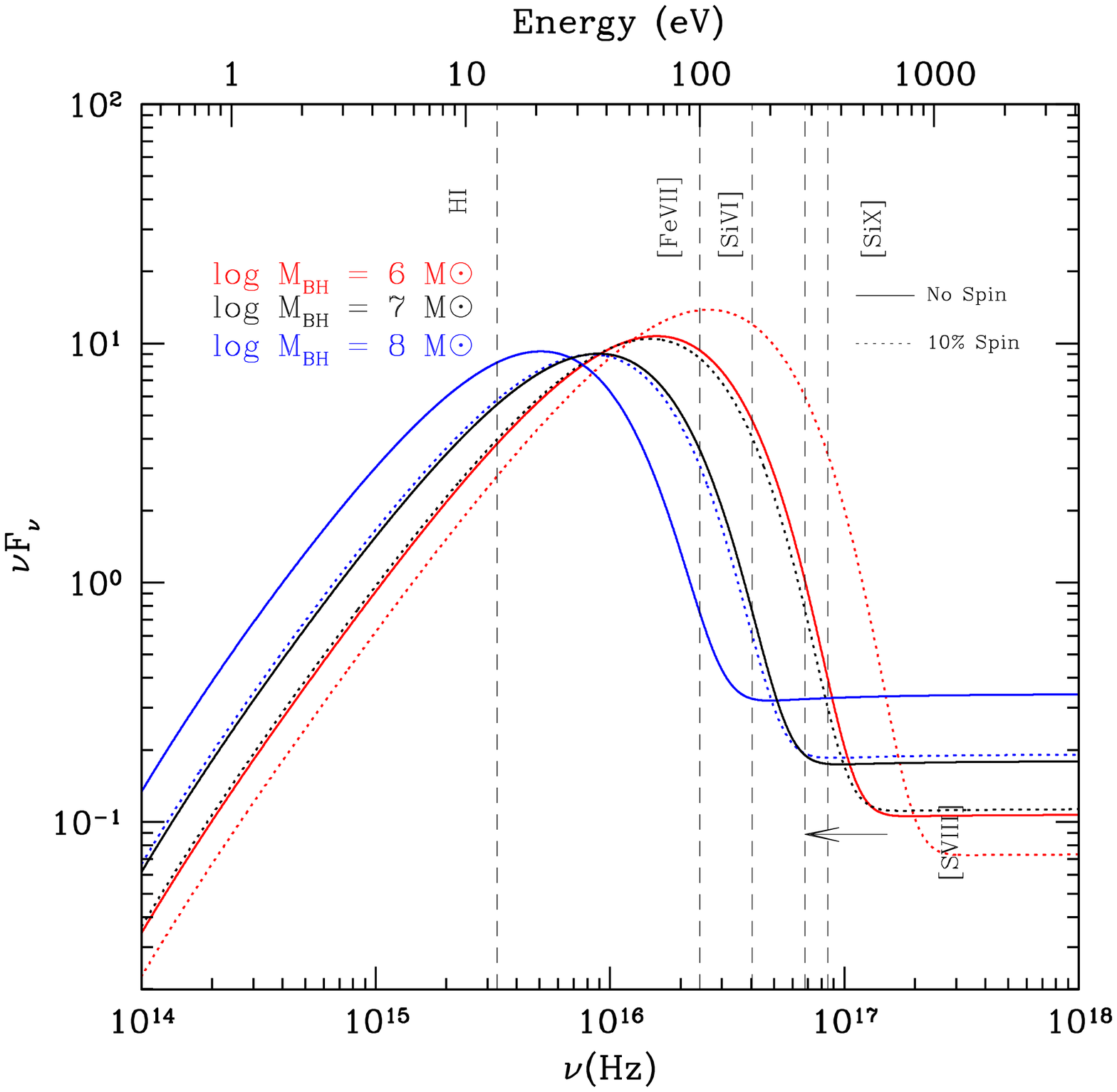}
    \caption{AGN ionizing continuum as per  Eqs. \ref{eq:sed} and \ref{eq:temp_disc} for three black hole masses for the cases of no spin (full lines) and spin of $10\%$ of the maximum spinning (dotted lines). The curves are normalised to \LLEdd{}=0.1. Vertical dashed lines mark the IPs of the lines used in the analysis.}
    \label{fig:continuum}
\end{figure}

If CL are photoionized by the accretion disc, following accretion  predictions, a  trend between CL strength and BH mass may be expected specifically for those lines whose IP get relatively close to the disc peak emission   (Fig.~\ref{fig:continuum}), i.e. a  potential  correlation between  CL strength  and the disc effective temperature, $T_{\rm{disc}}$ is foreseen. This possibility is investigated below.

Following on the  thin disc approximation, $T_{\rm disc}$ for a Schwarzschild BH can be approximated as e.g. \citep{lynden-bell69,frank+02}:

\begin{equation}
\begin{split}
 T_{\rm disc} = & 1.75 \times 10^5 K ~ \left(\frac{M_{BH}}{10^8 M_{\odot}}\right)^{1/4} \times \left(\frac{\left(\frac{dM}{dt}\right)}{0.1}\right)^{1/4}\\
 & \times \left(\frac{\eta}{0.06}\right)^{-1/4} \times \left(\frac{Rin_{G}}{6}\right)^{-3/4}
 \label{eq:tbbb}
\end{split}
\end{equation}

where, $M_{BH}$ is the BH mass, $(dM/dt)$ is the accretion rate in Eddington units, $\eta$ is the BH radiation efficiency, $Rin_{G}$ is the inner-most stable circular orbit  for a non-rotating  black hole in terms of the gravitational radius $R_G = G ~ M_{BH} ~/~ c^2$ (\textit{G} is the gravitational constant,  \textit{c} is the velocity of light). 
The equation is normalised to the average parameters characterizing the objects in the sample: average $M_{BH}$ of 10$^8$ M$_{\odot}$, average accretion rate $dM/dt$ $\sim$ 0.1 $(dM/dt)_{Edd}$, and accretion efficiency of 6\%.  

When a non-zero BH spin is assumed, a more realistic approach is obtained given the mounting evidence for spinning BHs \citep[see][and references therein]{lynden-bell69,campitiello+18,reynolds2019}. The inner-most stable orbit becomes smaller and  $T_{\rm disc}$ increases accordingly. In the case of co-rotation, the disc temperature reaches the highest values. Assuming a conservative spin of 10\%,  $ a = 0.1 ~G~M_{BH}/c^2$ (hereafter a= 0.1) and co-rotation yields $Rin_G = 2.5~ R_G$. Keeping the other parameters the same as above, and rounding $\eta$ to  $\eta = 10\%$ -  we get

\begin{equation}
 T_{\rm disc} = 3\times 10^5~ K \left(\frac{M_{BH}}{10^8 M_{\odot}}\right)^{-1/4}
 \label{eq:temp_disc}
\end{equation}

For the range of BH masses in this work $ 10^6 M_{\odot} < M_{BH} <~ 10^8 M_{\odot}$, and spin \textit{a} = 0.1, $T_{\rm disc}$ gets in the range of a few  $10^5 - 10^6$~K, decreasing with increasing BH mass. For higher spin, e.g. $a$ = 0.5, $T_{\rm disc}$ increases by factor of two.

Fig. ~\ref{fig:continuum} compares the ionizing continuum for the cases of non-spinning and spin $a=0.1$. It can be seen that only in the case of spinning BH the disc peak temperature moves close to the IP energies of [\ion{Si}{vi}] and [\ion{Fe}{vii}]. Yet, for the range of BH mass considered,  the accretion disc energies  fall short from those required to account for [\ion{Si}{x}] and [\ion{S}{viii}] lines. For the case of non-rotating BH, disc photons become overall less energetic for all the CLs.
 
 
On the above premises, for the range of BH masses in this work, a   trend between $T_{\rm disc}$ and the excitation state of [\ion{Si}{vi}] or [\ion{Fe}{vii}] lines may be expected, whereas no trend involving the higher IP lines [\ion{Si}{x}] or [\ion{S}{viii}] is foreseen. For the latter, they are however expected to be good proxies of the disc temperature for  BH mass below 10$^6$ M$_{\odot}$, as inferred from \citet{cann+18} simulations. We believe this prediction may be at the root of the observed [\ion{Si}{vi}] ~/~Br$\gamma_{\rm broad}$ -- $M_{BH}$ correlation, and  of the lack of any relation  with the higher IP lines \seight{} or \siten{} (Fig. 1). We test this hypothesis over the next subsection.

We note, however, that for the \feseven{} line, a weak trend with BH mass is observed,  (Fig. \ref{fig:ratios}), even though $Si^{5+}$ and $Fe^{6+}$
have similar IP. We believe the causes are that [\ion{Fe}{vii}] / H$\beta_{broad}$ is subjected  to  the large variability exhibited by H$\beta$ - variability studies  in the near-IR indicate the Paschen and Brackett  lines being less variable \citep{landt+11} - and the inherent uncertainty in the measurement of broad H$\beta$,  affected by a stronger underlying power-law continuum and the \ion{Fe}{ii} pseudo-continuum.

Fig. \ref{fig:continuum} also suggests a dependence as well  with  lower IP lines, e.g  [\ion{O}{iii}] $\lambda$5007 (IP = 35 eV). However the observational evidence for such trend would in general be elusive  as intermediate to low ionization lines, being formed at larger radii from the center, are subjected to additional formation mechanisms as star formation and shocks.

\subsection{Coronal Line emission as diagnostic of the disc temperature}
\label{modelling}

To probe the CL emission as a proxy of the disc temperature, we make use of the photoionization code \textmyfont{CLOUDY} \citep{ferland+17}. The  goal  is to test whether the observed linear trend  [\ion{Si}{vi}] /Br$\gamma_{\rm broad}$ -- BH mass and lack of correlation with the higher IP lines can be reproduced in CLOUDY. To that aim, a range of electron densities, $n_{e}$, cloud's distances, \textit{r}, and ionisation parameters, \textit{U}, that commensurate with the observed emissivity and location of CL emission  are probed. 

Specifically, the selected range of parameters are as follows.
The CL emission has been resolved, to extend over several tens of parsec to upmost 100 pc \citep{prieto+05}, the bulk of the emission is mostly nuclear and accounted by photoionization \citep[][]{rodriguez-ardila06,muller-sanchez+11,rodriguez+17,may18}.
 
As input to CLOUDY, the distance of the CL clouds to the centre \textit{r}, is taken in the   0.3~pc $\lesssim r \lesssim$ 100~pc range, the lowest limit set by the location of the  broad-line-region \citep{bentz13,martinez-aldama2019}.

Models are run for densities, $n_{e}$,  in the  $10^4 \leq n_{e} \leq 10^7$ cm$^{-3}$  range, the upper limit set by the critical density of the CLs probed - $n_{e} > 10^8$ cm$^{-3}$, the lower one by  the average densities inferred from the mid-IR [\ion{Ne}{v}] lines, IP=97 eV, $10^3 - 10^4 \rm{cm^{-3}}$ \citep{Fernandez-Ontiverosetal2016}, and ionization parameter, \textit{U}, in the high end, -1 $\leq \log U \leq$ 1 range, to cope with CL observed emissivity  \citep{ferguson+97, rodriguez-ardila06}.   

The input ionizing continuum is  Eq. \ref{eq:sed} (Fig.~\ref{fig:continuum}), normalised to \LLEdd{} = 0.1, representative of the AGN in the sample - average L$_{bol}$  $\sim 10^{44}$\,erg\,s$^{-1}$.  Solar abundances and a cloud size equivalent to a H column density, ${N_{H}}$ = 10$^{23}$\,cm$^{-2}$ are used.

\begin{figure*}
\gridline{\fig{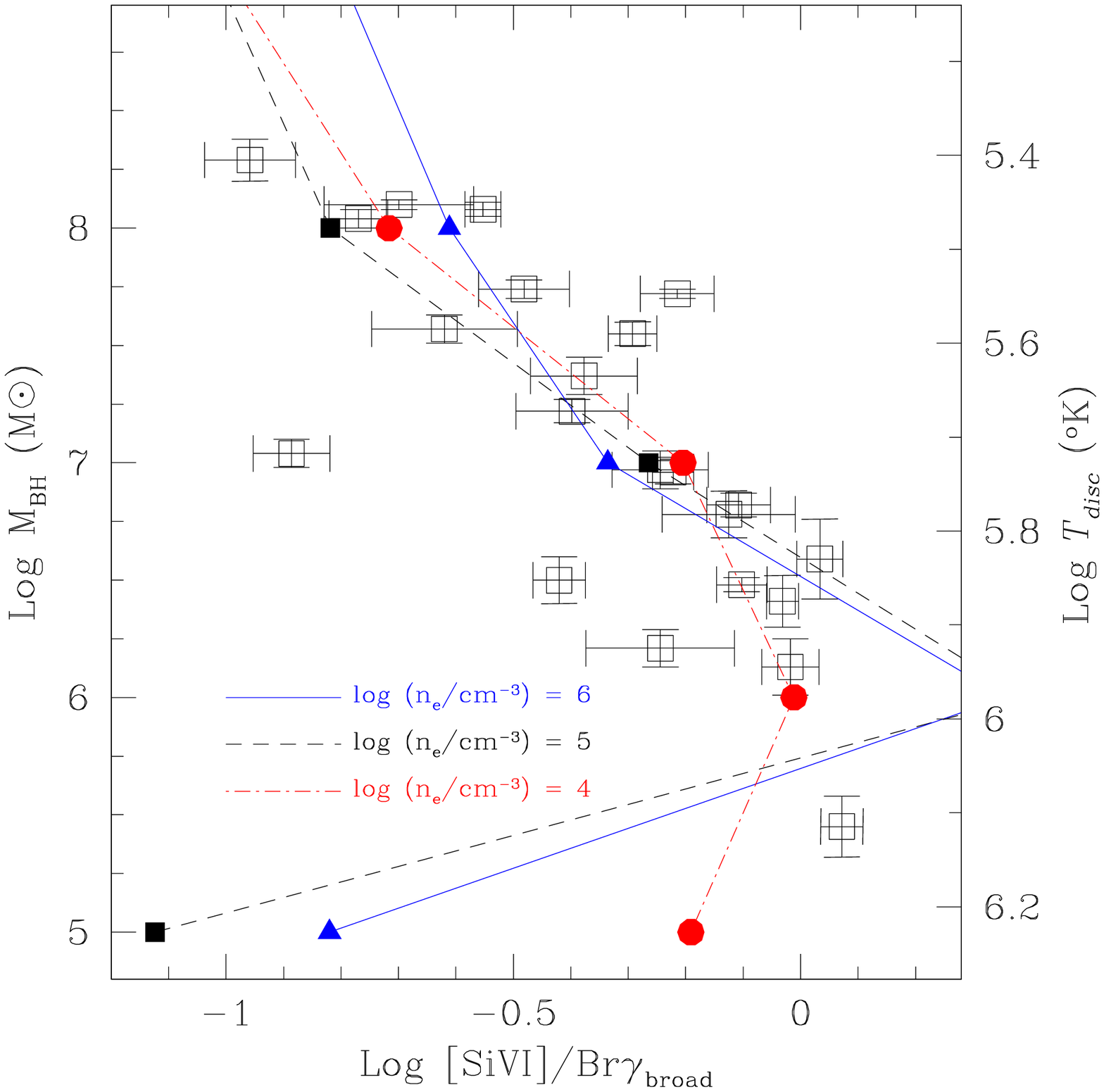}{0.5\textwidth}{(a)}
          \fig{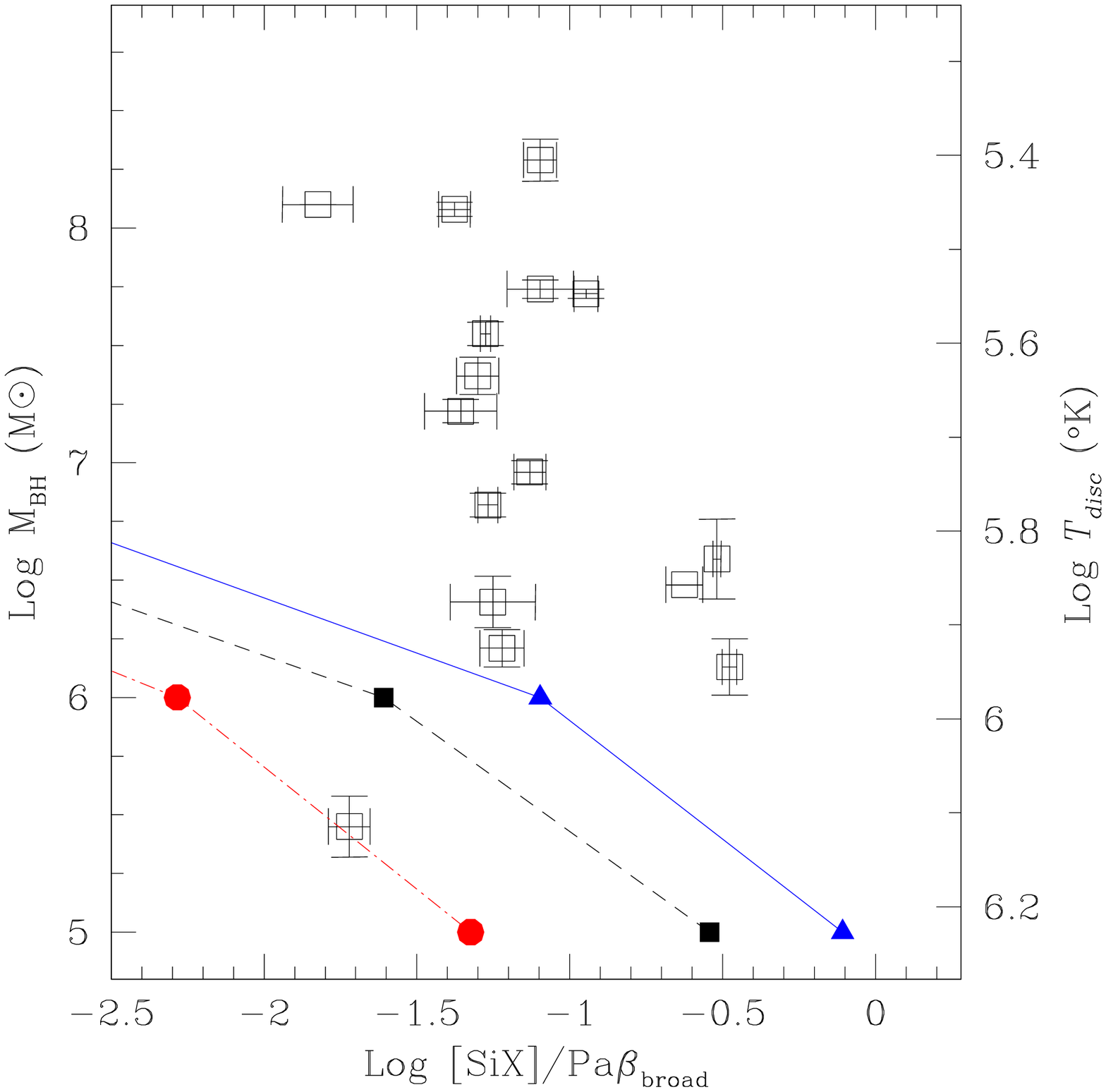}{0.5\textwidth}{(b)}}
\gridline{\fig{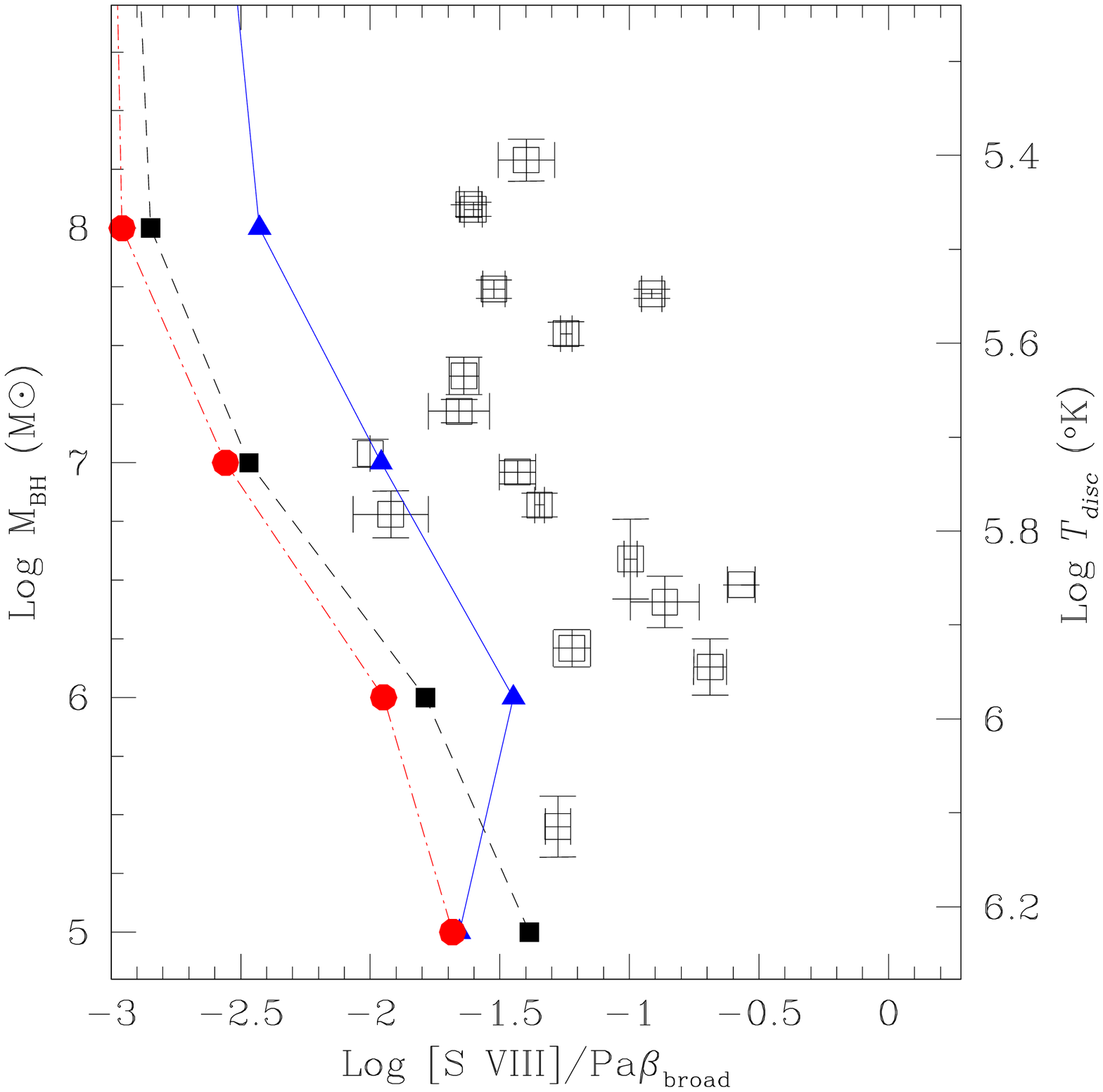}{0.5\textwidth}{(c)}
          \fig{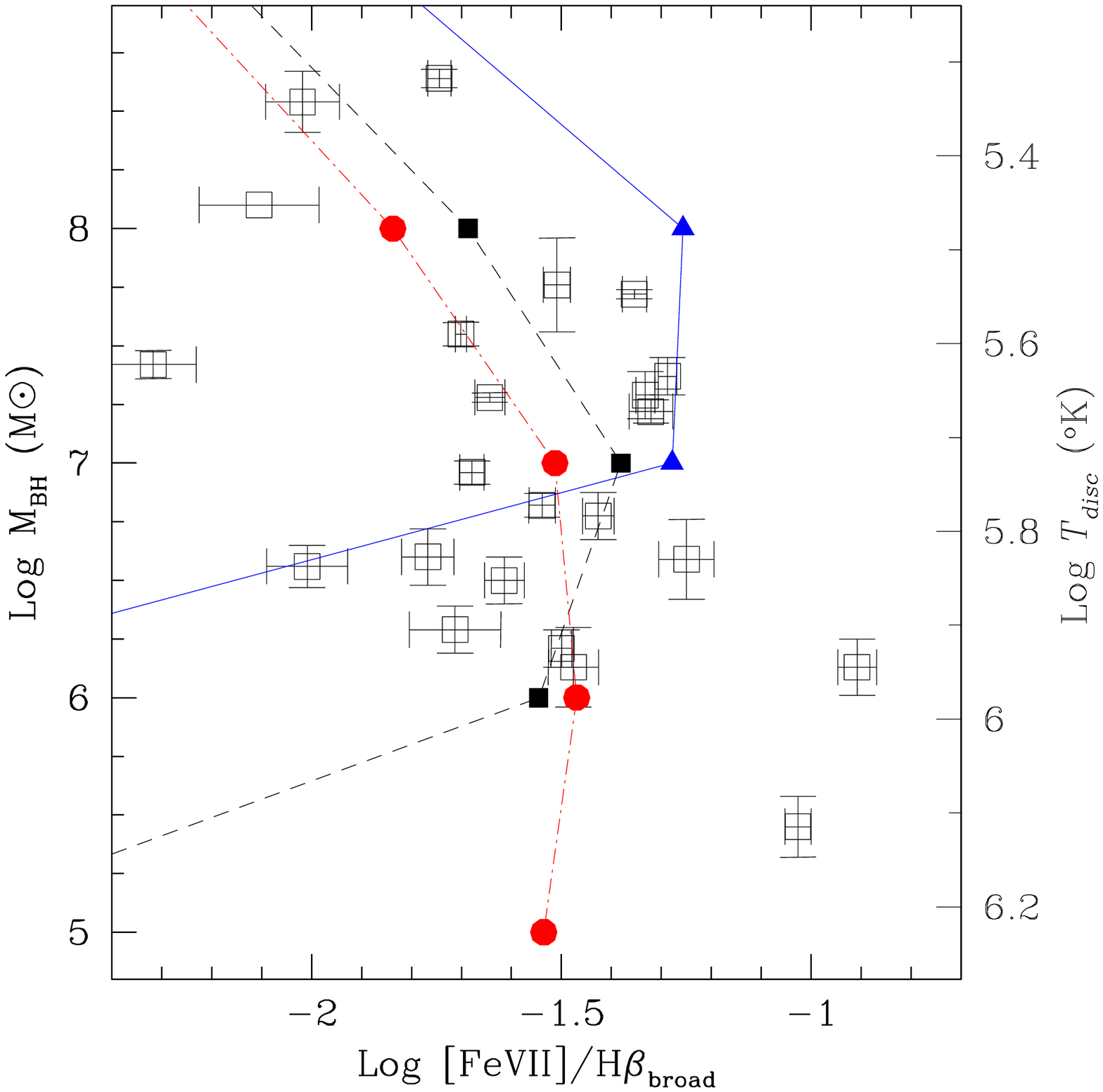}{0.5\textwidth}{(d)}}          
    \caption{ CL ratios predicted by \textmyfont{CLOUDY} for clouds of gas density log $n_{\rm H}$/cm$^{-3}$ = 6 (blue line), 5 (black line) and 4 (red line), located at a distance  to the centre, r=0.3 pc and log U=0.25,  as a function of  the accretion disc temperature as defined in Eq. \ref{eq:tbbb} for spin of 10\%,  in the right vertical axis, and as a function of BH mass, following Eq. \ref{eq:temp_disc}, in the left axis. Data points as in Fig. \ref{fig:ratios} are plot as open squares. Panel (a) shows the results for [\ion{Si}{vi}]/Br$\gamma$, panel (b) is for [\ion{Si}{x}]/Pa$\beta$, panel (c) plots [\ion{S}{viii}]/Pa$\beta$ and panel (d) is for [\ion{Fe}{vii}]/H$\beta$. 
    }
    \label{fig:models}
\end{figure*}

Fig.~\ref{fig:models} shows \textmyfont{CLOUDY} predictions for  the  CL ratios employed in this work  as a function of $T_{\rm disc}$ (in the right axis), and of $M_{BH}$ (on the left axis) following the transformation in Eq. \ref{eq:temp_disc}. Data are shown with open squares on top (same as in Fig. \ref{fig:ratios}). \textmyfont{CLOUDY} predictions that best constrain the data are shown. Limited cases are discussed in the text. We note that as the CL data is normalised to HI broad line region, to compare with \textmyfont{CLOUDY} predictions, these being  derived for densities at least two orders of magnitude below that of the broad line region, a correction factor to the models is applied. This factor is inferred from the compilation of narrow and broad H fluxes in a sample of 54 nearby AGN \citep{riffel+06}, the narrow/broad Br$\gamma$ ratio being in the 8 - 20 range. A factor 15 was chosen as  best compromise to scale down \textmyfont{CLOUDY}
predictions to the range of \sisix{}/Br$\gamma_{\rm{broad}}$ observed. Having fixed the scale for this ratio, those for the other CL line ratios were derived by imposing theoretical \ion{H}{i} recombination ratios. Accordingly, a factor of 25 is applied to \textmyfont{CLOUDY}'s [\ion{Fe}{vii}]/H$\beta$, and 90 to both [\ion{S}{viii}]/Pa$\beta$ and [\ion{Si}{x}]/Pa$\beta$.

Focusing  on [\ion{Si}{vi}]/Br$\gamma_{broad}$, \textmyfont{CLOUDY} results are similar in the  range of densities probed ($10^4 \leq n_{e} \leq  10^7$), enclosing all the observed data. Best predictions are for cloud  distances, $ r \leq 10$~pc, and $ -1 < {\rm log}~U < \sim 0.25$ and  $n_{e} \sim 10^5$ cm$^{-3}$, which give the closest to the linear dependence observed for the  BH mass range in this work, $ > 10^6 - 10^8$ M$_{\odot}$ (left axis). For BH mass below $10^6$ M$_{\odot}$, \textmyfont{CLOUDY} predictions change abruptly. In any case, only one galaxy falls in this range.

\textmyfont{CLOUDY} dependence with distance \textit{r} is that  of a slight increase in the CL ratios with increasing \textit{r}, model's trend are similar though.  \textit{r} $\sim$ 0.3 pc (used in Fig. \ref{fig:models}) provides best agreement. Predictions for clouds at tens or hundred of parsecs failed at reproducing the data unless densities are much smaller, but this would penalize the CL emission strength, making it undetectable \citep[e.g.][]{contini02}.
Regarding $U$, models with  $\textit{U} > 0.25$ progressively depart from the observed linear trend,  $\textit{U} < -1 $ severely penalises [\ion{S}{viii}] and [\ion{Si}{x}] emissivities.

Focusing on [\ion{Fe}{vii}]/H$\beta_{\rm{broad}}$,  the same  models that constrain \sisix{}/Br$\gamma_{\rm{broad}}$ also account for the spread of the iron ratio  (Fig. \ref{fig:ratios}). This consistency in the modeling  may \textit{a priori} be expected if considered that the IP of \sisix{} and \feseven{} are close to the peak temperature of the accretion disc for the range of BH mass considered (Fig. \ref{fig:continuum}), so both lines should be sensitive to the spectrum of the accretion disc although their dependence with the $T_{\rm disc}$ may be different. As discussed in former sections, a linear dependence with BH mass as that seen with \sisix{}/Br$\gamma_{\rm broad}$ was also expected for [\ion{Fe}{vii}] / $H\beta_{broad}$, which is not the case (Fig. \ref{fig:ratios}). Interestingly, \textmyfont{CLOUDY} does not predict a linear dependence with the temperature but a rather complicated one, which already sets aside this ratio as a temperature indicator.  A reason  may reside on the fact that collisional excitation of optical lines is more sensitive to the temperature than the IR lines.

None of the models above pass any close  to the observed  [\ion{Si}{x}]/Pa$\beta_{\rm{broad}}$ and [\ion{S}{viii}]/Pa$\beta_{\rm{broad}}$ values (Fig. \ref{fig:models}). Both line ratios show a scatter diagram when compared with BH mass (Fig.~\ref{fig:ratios}), which we attribute to their  IPs falling at the high energy end of the accretion disc spectrum (Fig.~\ref{fig:continuum}) for any of the BH mass considered. In this case, no dependence with $T_{\rm disc}$ is expected, which is hinted by  \textmyfont{CLOUDY} predictions. It can be seen in Fig.~\ref{fig:models} that the prediction for [\ion{S}{viii}]/Pa$\beta_{\rm{broad}}$ is a very weak dependence with the temperature, models are almost a straight line in $T_{\rm disc}$;  a stronger dependence is predicted for [\ion{Si}{x}]/Pa$\beta_{\rm{broad}}$ but  for  the high temperature range only. It should be noted that the range of parameters probed  are at the limit of what is feasible for coronal emission in terms of maximum density and minimum \textit{r}. Increasing the density, decreasing \textit{r}, will progressively shift the models to the loci of these data but will not change the dependence with $T_{\rm disc}$ - for the range of accretion discs -BH mass - considered.

\section{Overall view:  potential of coronal gas for  BH mass determination and  disc accretion  }

Using \textit{bona-fide} BH mass estimate from reverberation mapping and the line ratio [\ion{Si}{vi}]~1.963$\mu$m/Br$\gamma_{\rm broad}$ as  genuine tracer of the AGN ionizing continuum, a novel BH-mass scaling relation over almost three orders of magnitude in BH mass, $10^6 - 10^8$ M$_{\odot}$, is found (Fig. \ref{fig:ratios}). It follows the dependence $M_{BH} \propto$ ([\ion{Si}{vi}]/Pa$\beta_{\rm{broad}}$)$^{1.99 \pm 0.37}$, with a dispersion in BH mass of 0.44 dex (Sec. \ref{sec3}). Following on the thin accretion disc approximation and after surveying a basic parameter space for CL production, we believe that a key parameter driving this correlation is the effective temperature of the accretion disc, the observed correlation being  formally  in line with the thin disc prediction  $T_{\rm{disc}} \propto {M_{\rm BH}}^{-1/4}$.

For a suitable range of densities $n_{\rm e} ~ \sim 10^{4 - 6}$~cm$^{-3}$, and  ionization parameter to account for CL emissivities, photoionization models yield a narrow dependence  log($T_{\rm{disc}}$) -- log([\ion{Si}{vi}]/Br$\gamma$)  close to the linear trend observed between log$(M_{BH})$ and log([\ion{Si}{vi}]/Br$\gamma_{broad}$).
On the assumption that the difference between the broad (used in Fig.~\ref{fig:ratios}) and narrow (used in \textmyfont{CLOUDY})  $Br\gamma$  flux is a  scale factor of the order of 15, as inferred from observations (Sec. \ref{modelling}),  \textmyfont{CLOUDY} predictions  fall onto the empirical BH mass - ([\ion{Si}{vi}]/Br$\gamma_{\rm broad})$ relation.

The  BH mass - CL dependence  is  sensitive to the IP of the CL employed. The use of [\ion{Si}{vi}]~1.963~$\mu$m restricts the dependence to BH masses in the range of  $10^6 $ to  $10^8$~M$_{\odot}$ (Fig. \ref{fig:ratios}), potentially up to  $10^9$~M$_{\odot}$ but not yet tested, presumably because of the disc temperature for these masses favors [\ion{Si}{vi}] emission. In the same line of reasoning, an equivalent relationship involving  other CLs, e.g, [\ion{Fe}{vii}]~$\lambda$6087, with IP  close to the peak emission of the disc for these range of BH mass should  be expected. The complex dependence of [\ion{Fe}{vii}] / H$\beta_{broad}$ with temperature as predicted by \textmyfont{CLOUDY}  makes this line unsuitable for $T_{\rm disc}$ and, in turn, for BH mass diagnosis (Sect. \ref{accretion-disc}). Other prominent CLs are those from   Ne$^{4+}$ IP = 97 eV. Yet, the ones in the UV are much subjected to reddening. Those in the mid-IR  are available for few sources  with good BH mass determination. Still, we advance from first analysis a positive result.

Conversely, for the same BH mass range and objects above, no correlation involving the higher IP lines, [\ion{Si}{x}] and [\ion{S}{viii}], IP $> $250 ~eV is observed (Fig. \ref{fig:ratios}). This is because the IP of these lines sample hotter discs, and in turn smaller BH masses. Accordingly, no dependence with $T\rm{_{disc}}$, within the range of BH mass under study, is foreseen, which is indeed consistent with \textmyfont{CLOUDY} predictions (Fig. \ref{fig:models}). These lines are however expected to show a dependence with the disc temperature with decreasing BH masses below 10$^6$ M$_{\odot}$, becoming thus  potential candidates for BH mass determination at intermediate BH masses, as suggested by \citep{cann+18}. Testing this low mass end is currently limited by the lack of suitable data.

Above $10^8$ M$_{\odot}$,  disc temperatures are foreseen in the $10^5$~K regime.  
Lower ionization lines would then be more favoured than higher ones. As expected, \textmyfont{CLOUDY} predicts a general decrease in [\ion{Si}{vi}]/Br$\gamma$ with decreasing temperature (Fig.~\ref{fig:models}); observationally, [\ion{Si}{vi}]/Br$\gamma_{broad}$ decreases with increasing BH mass (Fig. \ref{fig:ratios}). The high BH mass tail in the local universe is the realm of elliptical and bulge dominated objects often associated with LINERS. Parsecs-scale near-IR observations have so far proven the elusiveness of CL emission in a few nearby cases \citep{muller-sanchez+13}, in line with this prediction. These sources would appear as upper limits at the high mass range in Fig.~\ref{fig:ratios}.

If the above predictions are correct, some intrinsic scatter in the proposed BH mass scaling relation should  be present. This may reflect a range of  properties of the individual objects. Prime assumptions involved in the modeling are  the radiation efficiency, accretion rate, and spin, all being fixed to a reference of 10\%  in the BH mass -- $T\rm{_{disc}}$ thin disc approximation relation (Eq. \ref{eq:temp_disc}). Among these, an increase in spin or accretion rate translates into a progressive increase in $T\rm{_{disc}}$ \citep{campitiello+18,campitiello+19}. An increase in efficiency would produce the opposite effect. Because of the exponential dependence of these parameters, the expected increase in efficiency with increasing spin would yield little change in $T\rm{_{disc}}$. An increase of at most a factor two in $T\rm{_{disc}}$ is achieved for spin  0.5\%,  co-rotation and efficiency 10\%, but a factor of 1.4 for maximum spin, and  40\% efficiency. Overall the net effect in BH mass is in the 10 dex  range. Regarding \textmyfont{CLOUDY} predictions, the dependence of [\ion{Si}{vi}]/Br$\gamma$ on density and ionization parameter is small (Fig. \ref{fig:models}) provided we are in the optimal range for production of CL emission, $n_e > 10^4$~cm$^{-3}$, log~$U > -1 $. With a final compendium of 31 objects, the  dispersion in BH mass with the proposed  calibration is  0.47 dex - one $\sigma$,  which compares with  0.44 dex inferred  from the $M-\sigma$ relation in 49 galactic bulges with direct dynamical BH mass estimate \citep{gultekin+06}.

The present BH mass scaling relation is restricted to Type~1 AGN including narrow line Type 1, and BH masses in the $10^6~-~10^8~$ M$_{\odot}$ range. The limitation to Type~I is currently driven by the need to normalise to broad \ion{H}{i} gas. Attempts to normalise to narrow \ion{H}{i} introduce excessive scatter possibly because of stellar contamination  and the large size of the \ion{H}{ii} region as compared with nuclear, parsec scale, CL region.

The proposed scaling offers an economic, physically motivated, alternative for BH estimate using single epoch spectra that avoid large telescope time (reverberation mapping) or absolute flux calibration (the continuum luminosity method, \citet{landt+13} and references therein). With James Webb Space Telescope and big surveys in the IR region, large samples of AGNs could be weighted using this  approach.

\section*{Acknowledgements}
We are grateful to H. Netzer for critical review,  E. Churazov, B. Czerny, A. Askar for discussions.
A.R.A acknowledges partial support from CNPq Fellowship (311935/2015-0 \& 203746/2017-1), S.P. from the Polish Funding Agency National Science Centre, project 2017/26/\-A/ST9/\-00756 (MAESTRO  9) and MNiSW grant DIR/WK/2018/12.  S.P would like to acknowledge the computational facility at Nicolaus Copernicus Astronomical Center.





\bibliography{article}

\appendix
\section{Optical Spectra}
\restartappendixnumbering


 \begin{figure}[hbt!]
	\includegraphics[width=\columnwidth]{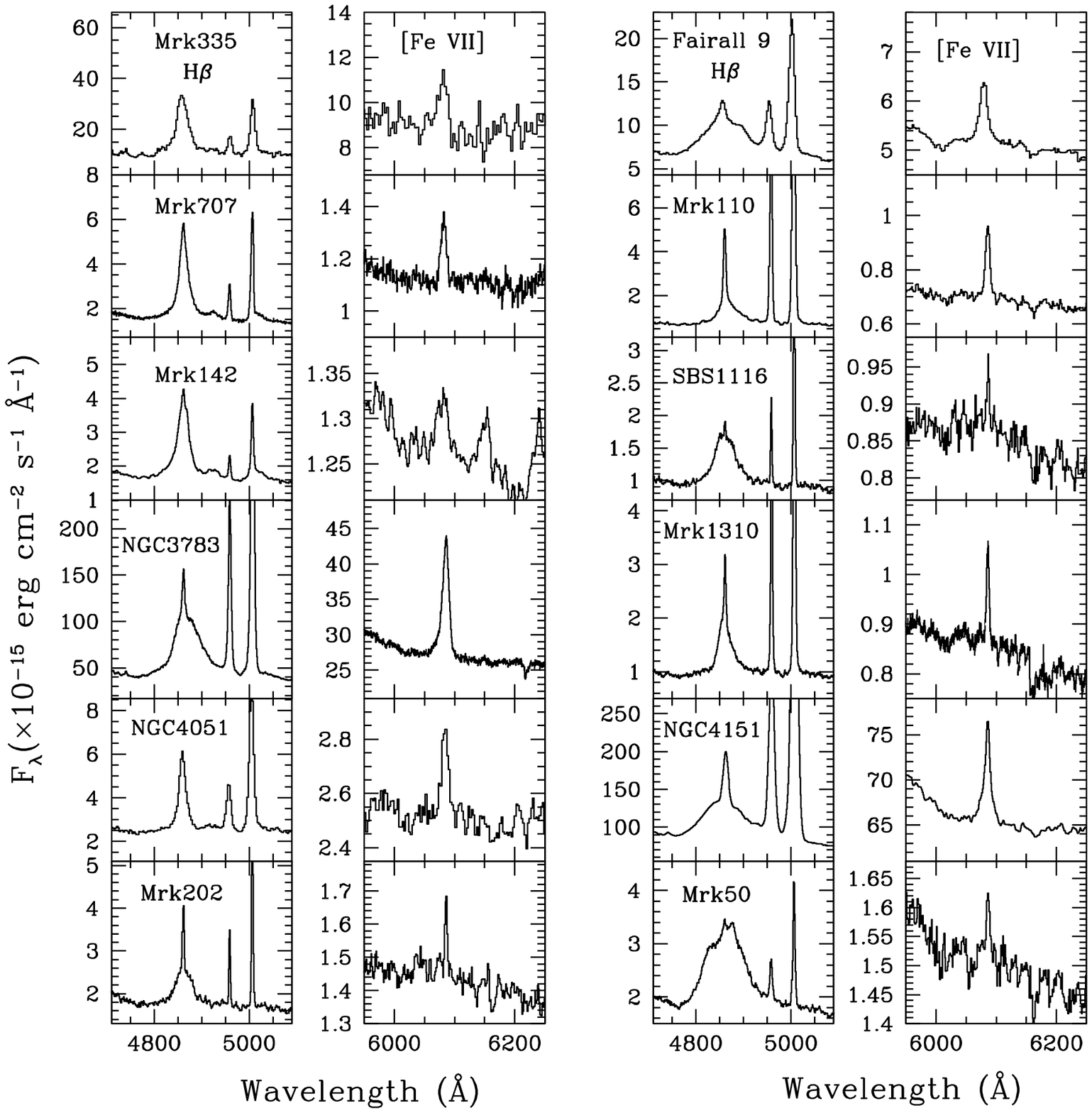}
    \caption{Optical Spectra of the AGN sample in rest wavelength. For each galaxy, the left panel shows the H$\beta$ line while the right panel shows the [\ion{Fe}{vii}]~$\lambda$6087 line. }
    \label{fig:optspec1}
\end{figure}

\begin{figure}
	\includegraphics[width=\columnwidth]{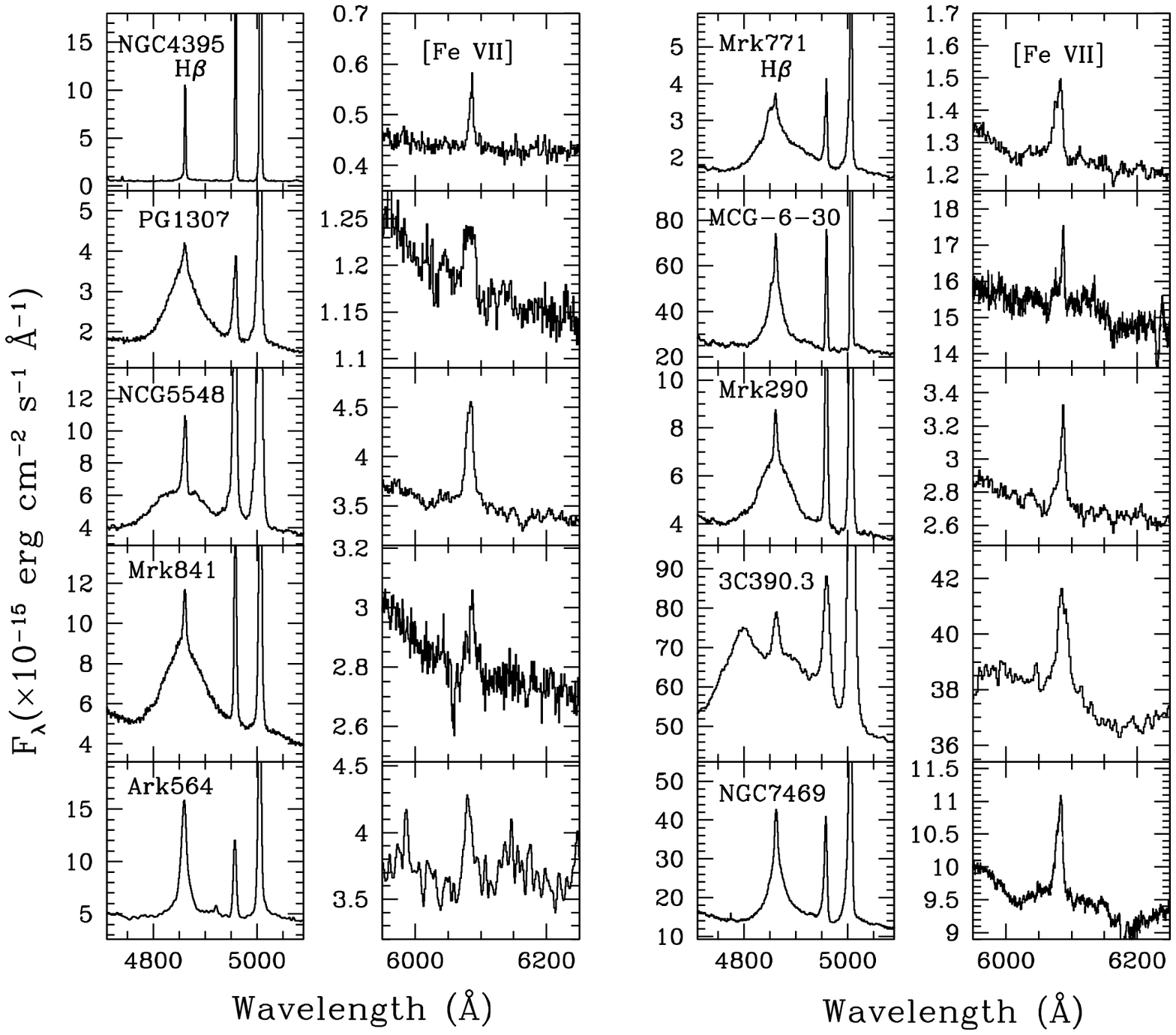}
    \caption{Cont. Fig.~\ref{fig:optspec1}. }
    \label{fig:optspec2}
\end{figure}

\begin{figure}
	\includegraphics[width=\columnwidth]{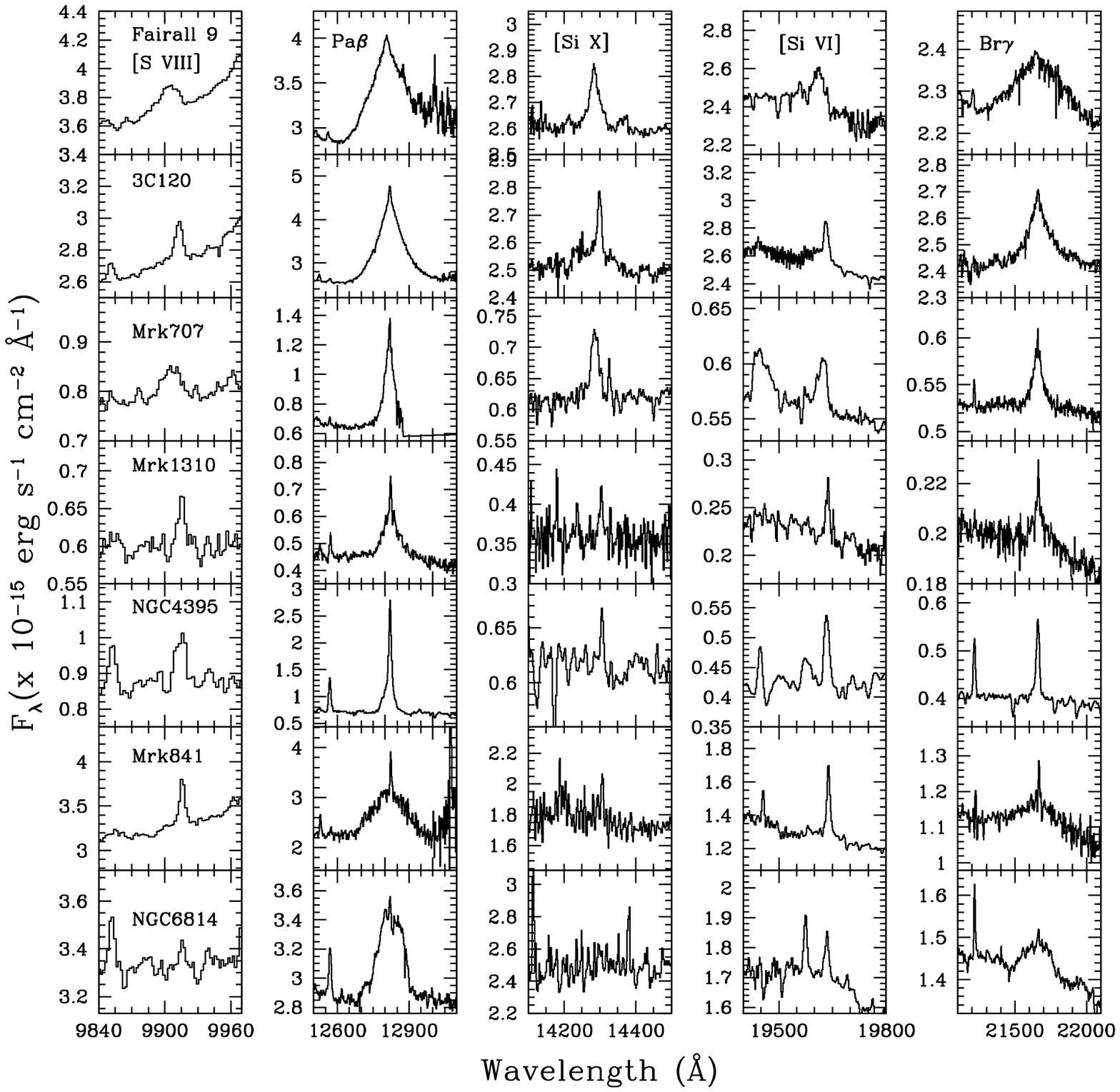}
    \caption{NIR spectra of the AGN in rest wavelength with no previous information in the literature. Each row presents the NIR lines relevant to this work detected in the sample. }
    \label{fig:nirspec1}
\end{figure}

\begin{deluxetable}{lccc}
	\tablecaption{Measured Fluxes, in units of 10$^{-15}$~erg~cm$^{-2}$~s$^{-1}$, of the broad \ion{H}{i} lines for the galaxy sample.
	\label{tab:fluxes}}
	\tablewidth{0 pt}
	\tablehead{
	\colhead{Galaxy} & 
        \colhead{H$\beta$} &
        \colhead{Pa$\beta$} &
	\colhead{Br$\gamma$}} 
	\startdata
Mrk\,335 &  115.4$\pm$1.2 & 170$\pm$5 & 26.7$\pm$3.1 \\
Fairall~9 & 49.0$\pm$0.3 & 107.9$\pm$8.4 & 43.1$\pm$4.7 \\
NGC\,863 &  ... & 49$\pm$3 & 10.4$\pm$2.9 \\
3C\,120   & ... & 223.0$\pm$7.0 & 46.3$\pm$6.5 \\
Mrk\,707 & 10.7$\pm$0.3 & ... & 4.8$\pm$0.4 \\
Mrk\,110  & 5.2$\pm$0.1 & ... & ... \\
NGC\,3227 &  ... & 168.7$\pm$10.2 & 20.0$\pm$3.9 \\
Mrk\,142 & 8.6$\pm$0.07 & ... & ... \\
SBS\,1116+583A	& 4.7$\pm$0.2 & ... & ... \\
PG\,1126-041 & ... & 101.8$\pm$3.7 & 13.3$\pm$0.07 \\
NGC\,3783 & 469.3$\pm$15.7 & 348.7$\pm$19.1 & 60.4$\pm$9.8 \\
Mrk\,1310 & 3.49$\pm$0.07 & 11.4$\pm$1.1 & 1.4$\pm$0.4 \\ 
NGC\,4051 &  4.8$\pm$0.2 & 66.6$\pm$1.7 & 13.1$\pm$0.8 \\
NGC\,4151 &  758.5$\pm$7.9 & 712.5$\pm$8.8 & 125.0$\pm$10.8 \\
Mrk\,202 & 6.30$\pm$0.17 & ... & ... \\
Mrk\,766 & 82.9$\pm$3.0 & 117.8$\pm$1.8 & 20.0$\pm$2.26 \\
Mrk\,50 & 16.0$\pm$0.5 & ... & ... \\
NGC\,4395 & 1.10$\pm$0.02 & 30.5$\pm$1.1 & 1.9$\pm$0.2 \\
Mrk\,771 & 12.5$\pm$0.2 & ... & ... \\
NGC\,4748 &  ... & 60.1$\pm$2.2 & 9.7$\pm$0.5 \\
PG\,1307+085 & 17.2$\pm$0.2 & ... & ... \\
MGC-6-30-15 &	110.0$\pm$0.4 & ... & ... \\
NGC\,5548 & 31.4$\pm$1.0 & 49.3$\pm$2.9 & 16.3$\pm$2.0 \\
PG1448+273 &  ... & ...  & 2.40$\pm$0.10 \\
Mrk\,290 & 23.9$\pm$0.5 & ... & ... \\
Mrk\,841 & 43.6$\pm$1.2 & 146.7$\pm$10.4 & 26.1$\pm$7.5 \\
3C\,390.3  & 43.3$\pm$0.7 & ... & ... \\
NGC\,6814 & ... & 81.0$\pm$13.3  & 11.9$\pm$3.5 \\
Mrk\,509 &  ... & 1824.7$\pm$77.9 & 349.0$\pm$21.8 \\
Ark\,564 & 16.0$\pm$0.2 & 59.0$\pm$1.5 & 5.7$\pm$0.4 \\
NGC\,7469 & 85.5$\pm$1.2 & 153.1$\pm$7.3 & 20.7$\pm$1.4 \\
\enddata
\end{deluxetable}

\end{document}